\documentclass{webofc}
\usepackage[varg]{txfonts}   
\usepackage{graphicx}
\usepackage{fancyhdr}
\usepackage{appendix} 
\begin{document}
\title{The impact of applying WildCards to disabled modules for FTK pattern banks on efficiency and data flow}

\author{\firstname{Khalil} \lastname{Bouaouda}\inst{1}\fnsep\thanks{\email{khalil.bouaouda@cern.ch}} \and
        \firstname{Stefan} \lastname{Schmitt}\inst{2}\fnsep\thanks{\email{sschmitt@mail.desy.de}} \and
        \firstname{Driss} \lastname{Benchekroun}\inst{1}\fnsep\thanks{\email{driss.benchekroun@cern.ch}}
        on behalf of the ATLAS collaboration.
}

\institute{Laboratory of High Energy Physics and Condensed Matter (LPHEMaC), Department of Physics, Faculty of Sciences A\"in Chock, Hassan II University, B.P. 5366 Maarif, Casablanca 20100, Morocco. 
\and
            Deutsches Elektronen-Synchrotron DESY, Hamburg
          }

\abstract{%
Online selection is an essential step to collect the most relevant
collisions from the very large number of collisions inside the ATLAS
detector at the Large Hadron Collider (LHC).
The Fast TracKer (FTK) is a hardware based track finder, built to
greatly improve the ATLAS trigger system capabilities for identifying
interesting physics processes through track-based signatures.
The FTK is reconstructing after each Level-1 trigger all tracks with
$ p_T>1 $ GeV, such that the high-level trigger system gains access to
track information at an early stage.
FTK track reconstruction starts with a pattern recognition step.
Patterns are found with hits in seven out of eight possible detector layers.

Disabled detector modules, as often encountered during LHC operation,
lead to efficiency losses. To recover efficiency, WildCard 
(WC) algorithms are implemented in the FTK system.
The WC algorithm recovers inefficiency but also causes high combinatorial
background and thus increased data volumes in the FTK system, possibly
exceeding hardware limitations.
To overcome this, a refined algorithm to select patterns is developed and
investigated in this article.}
\maketitle
\let\thefootnote\relax\footnote{Copyright 2018 CERN for the benefit of the ATLAS Collaboration.}
\let\thefootnote\relax\footnote{CC-BY-4.0 license}
\section{Introduction}
\label{intro}

\begin{figure}[h]
  \centering
  \sidecaption
  \includegraphics[width=0.6\textwidth]{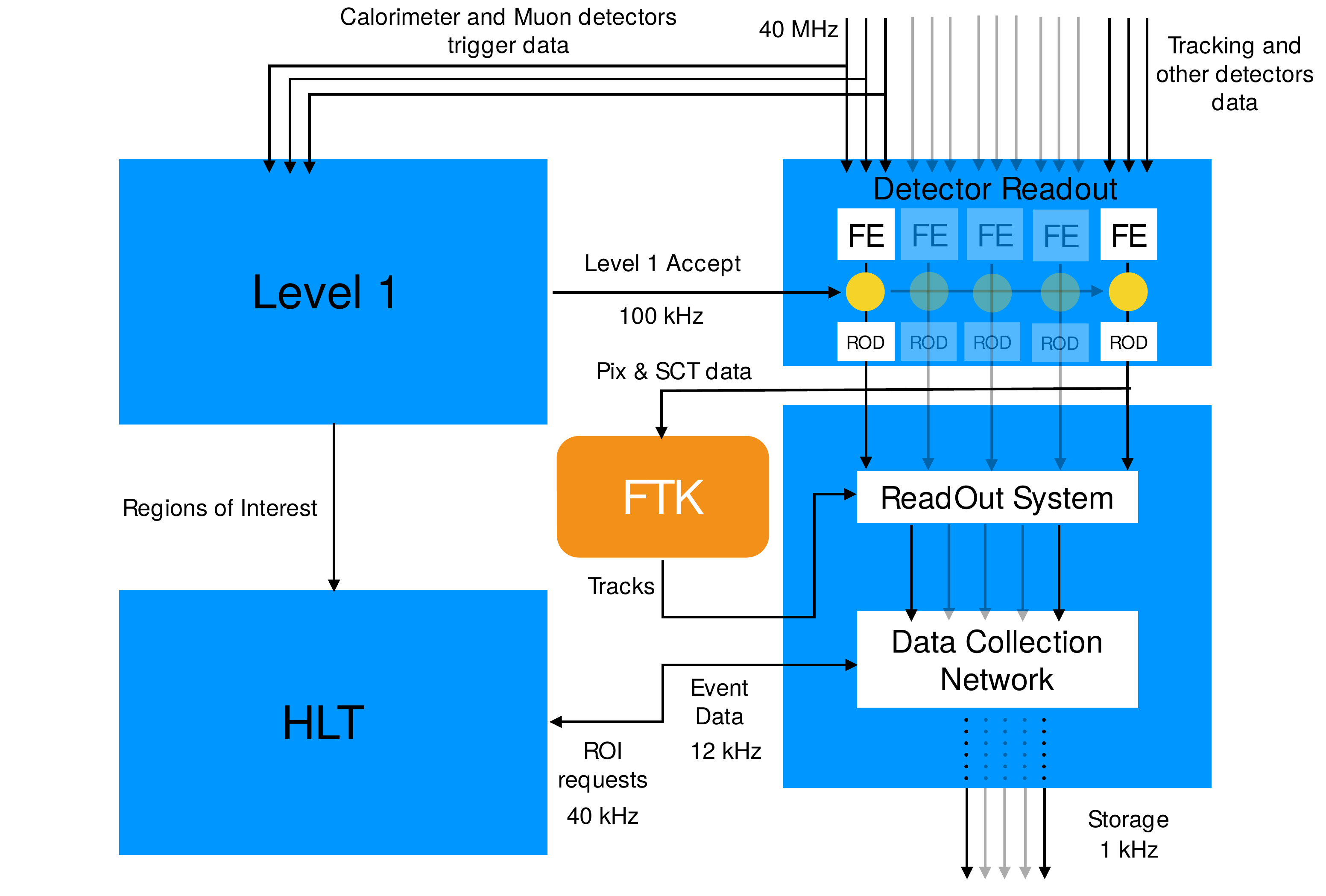}
  \caption{ATLAS trigger system and integration of the FTK.
    Given a Level-1 accept signal, event data are transferred from the
    front-ends (FE) to the readout drivers (ROD).
    The RODs send the data to large buffers (ReadOut System). Tracker
    data are also sent to the FTK system. The reconstructed FTK tracks
    are sent to the ReadOut System. The high-level trigger (HLT)
    requests event fragments, possibly including FTK tracks, from the
    ReadOut System and evaluates the final trigger decision.
    \label{fig:1}
    }
\end{figure}

The ATLAS trigger system \cite{HLT2016} is a combination of a
hardware-based Level 1 and software-based High Level Trigger (HLT),
which reduces the event rate from 40 MHz to an average
output rate of 1 kHz.
The Fast TracKer (FTK) \cite{Annovi2013} is a track finding system for
use with the ATLAS trigger, that finds and reconstructs tracks for
each event that passes the Level 1 trigger. 
Tracks are reconstructed with an average latency of $100\,\mu s$ at an
event input rate of $100$~kHz.
The FTK system uses data from the pixel detector and the semiconductor tracker
(SCT) and is designed to operate at instantaneous luminosities up to
$3\times10^{34}$~cm$^{-2}$~s$^{-1}$.
After processing, FTK provides the helix parameters and
hits for all tracks with $p_T>1$~GeV to the HLT.
The dataflow in the ATLAS
trigger system including the FTK is shown in Figure~\ref{fig:1}.

Handling the required amount of data is a serious design challenge.
To deal with the large input data volume and rate, FTK uses a highly
parallelised system, divided into 64 independent regions. The division
is made in 16 azimuthal segments, each of which is further divided into 4
segments in polar angle.
As of mid 2018, a partial FTK system is operational, covering a few
selected regions. Hardware installation and commissioning of the
full system is ongoing.

During operation with the LHC, tracking detectors may encounter
problems, which can lead to disabled detector modules.
These do not provide useful hit information and cause inefficiencies in
the FTK track reconstruction.
To recover efficiency, a WildCard (WC) algorithm is implemented in the
FTK.
Disabled modules on which WCs are set, are treated as if all their
channels were on for each event.
This does recover efficiency losses, but leads to a sizable increase in
the number of fake track segments, which has the potential to slow down or even
saturate the FTK system.
To control these effects, modifications to the pattern selection
scheme are implemented that reduce the amount of data while keeping a
reasonable track reconstruction efficiency.

The FTK system design is described in section \ref{FTKsys}.
The pattern selection is presented in section \ref{patbank}.
The effect of disabled modules and details of the algorithm to control
the number of fake track segments are described in section \ref{WC}.
The results are discussed in section \ref{results}.

\vskip-20pt
\section{FTK system design}
\label{FTKsys}
\begin{figure}[h]
	\centering
	\sidecaption
	\includegraphics[width=0.6\textwidth]{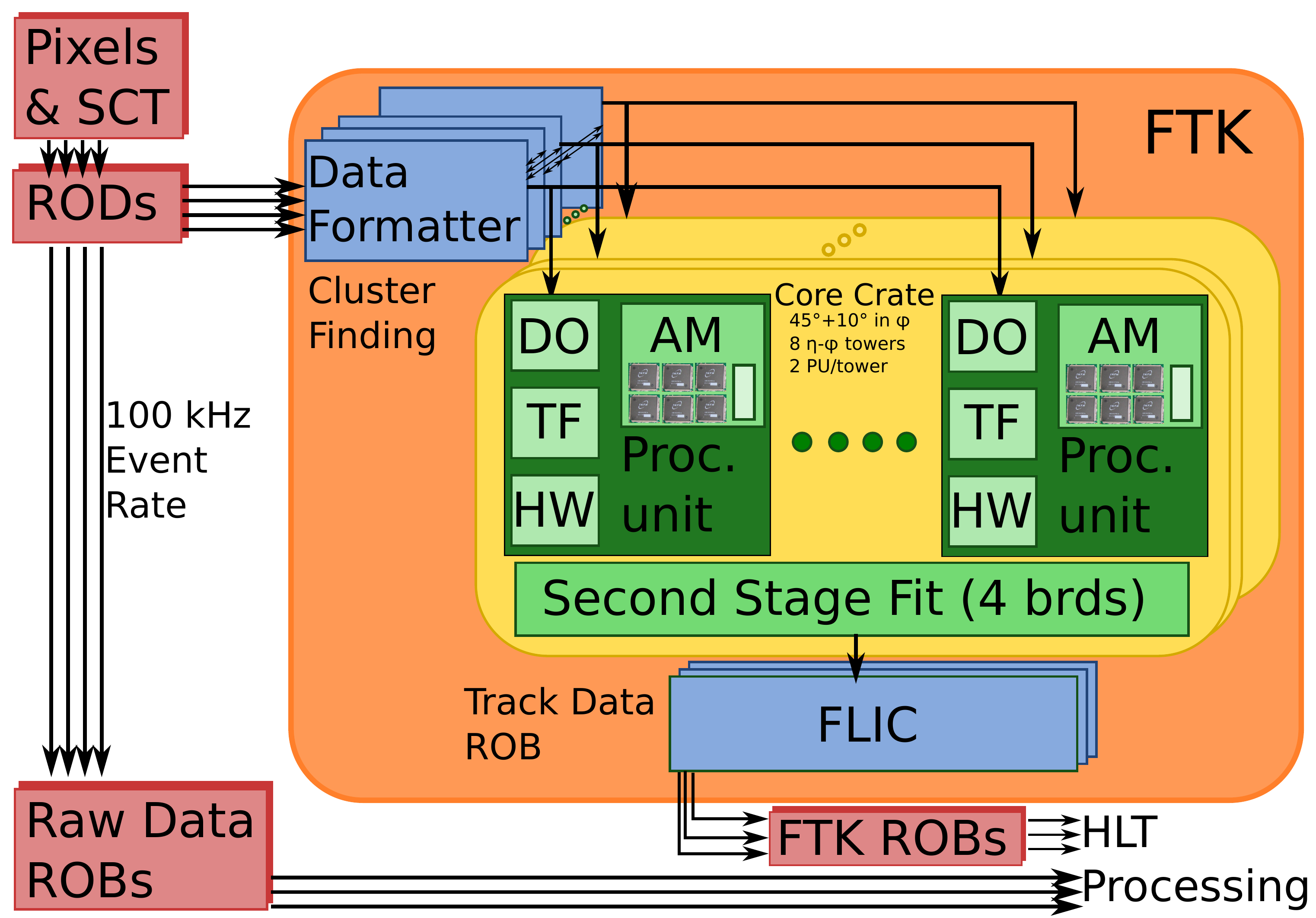}
	\caption{Functional sketch of the FTK system. The different components, namely
          the Data Formatter
          (DF), Data Organizer (DO), Associative Memory (AM) Track
          Fitter (TF), Hit Warrior (HW), Second Stage Fitt (SSB)
          and FTK Level 2 Interface Card (FLIC), are described in the text.}
	\label{fig:2}
\end{figure}

The FTK system is schematically shown in Figure \ref{fig:2}. Briefly,
the individual components take the following roles:
\begin{itemize}  \setlength{\itemsep}{-0.4\baselineskip}
\item The pixel and strip data are transmitted from the Read Out
  Drivers (ROD) to the Data Formatters (DF). 
\item DF mezzanine cards perform a clustering to form hits, in two dimensions
  for the pixel layers and in one dimension for the SCT.
  The DF distributes the hits according to the FTK segmentation into
  64 regions.
\item The Associative Memories boards (AM) compare hits to a total of
  $10^9$ predefined track patterns at coarse resolution and return a
  list of patterns with seven out of eight possible matches.
  Only eight of the 12 available silicon layers are used for pattern
  recognition.
  These are the three outermost pixel layers, the four axial SCT
  layers and one of the stereo SCT layers.
\item The Data Organizers (DO) are smart databases which receive the
  pattern or ``road'' numbers from the AM and send the corresponding
  hit information at full resolution to the Track Fitter (TF).
\item The TF determines helix parameters for track segments, each
  with up to 8 hits, using a linearized fit as described below.
  The track segments can be fitted at a rate of $10^9$ per second,
  using highly parallel processing. 
\item Good track segments are selected using certain quality criteria.
  Duplicate track segment removal is carried out by the Hit
  Warrior (HW).
\item The Second Stage Boards (SSB) perform both an extrapolation of the
  track segments to the remaining 4 layers and the final track fits
  using up to 12 hits.
  Again, good tracks are selected using certain quality criteria and
  duplicate tracks are filtered.
\item The FTK Level 2 Interface Card (FLIC) collects the tracks and
  sends them to the Read Out Buffers (ROB).
\end{itemize}

\subsection{Pattern recognition}
\label{patreco}

\begin{figure}[b]
	\centering
	\includegraphics[width=0.8\textwidth]{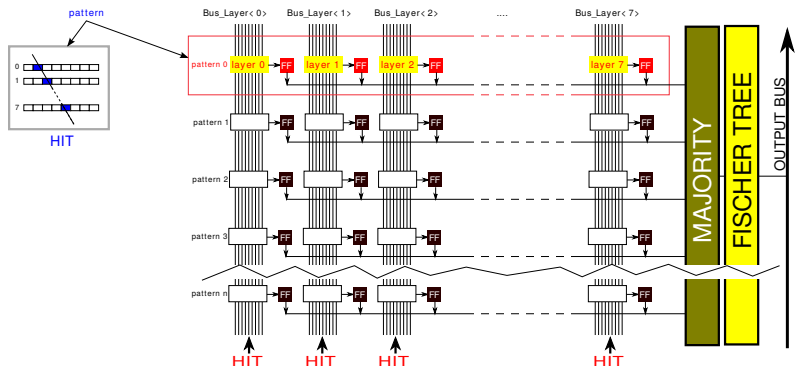}
	\caption{Schematic view of the associative memory chip logic.}
	\label{fig:3}
\end{figure}

The pattern recognition is the heart of the FTK system, using a dedicated
technology based on Content Addressable Memories (CAM), the AM chip
\cite{AM2017}. There are 64 AM chips per AM board and a total of 128
AM boards in the full FTK system.
Per chip, there are eight CAMS (one per detector layer), with 128K
addresses each. Prior to data taking, the CAMs are loaded with
predefined track patterns (pattern bank).
The CAM data lines comprise 12 ordinary bits and three bits with
ternary logic.
Data (coarse resolution hit positions) arrive through 8 independent
buses corresponding to the 8 detector layers. CAM matches are stored
in Flip-Flops (FF), one per CAM and per address. A majority logic 
detects addresses where 7 or 8 CAMs have a match. The resulting
addresses (road or pattern numbers) are read out using a Fisher tree.
A schematic view of the AM chip is shown in Figure \ref{fig:3}.

\subsection{Track fitting}
\label{trackfit}

Track fits or track segment fits are based on a linearized model,
where the helix parameters ($d_{0}$, $z_{0}$, $\eta$, $ \phi$ and
$q/p_{T}$) are calculated as \cite{Annovi2013}:
\begin{equation}
 p_{i}= \sum_{j} c_{ij} . x_{j} \ + \ q_{i}.
\end{equation}
The $ p_{i}$ are the tracks parameters, the $x_{j}$ are the hit
coordinates.
There are two coordinates for pixel layers and one per SCT layer.
The $c_{ij}$ and $q_{i}$ are ``fit constants'' which are valid in
small geometrical zones called ``sectors``.
The fit quality $\chi^2$ is also determined using another set of constants
and a quadratic form.
A FTK sector consists of a group of detector modules, one
module per layer. Dedicated sets of fit constants are determined for
each sector.

The full list of sectors with the corresponding fit constants is
determined by processing about $10^9$ muon tracks with the
ATLAS tracker simulation.
The muon tracks are drawn from uniform distributions  in the five track
parameters, $d_{0} \in [-2.2, 2.2]$~mm, $z_{0} \in [-120, 120]$~mm,
$|\eta| < 3$, $ \phi \in [-\pi, \pi]$ and $1/p_{T} \in [-0.8, 0.8]$~GeV$^{-1}$.
A single muon crossing a group of modules already defines a sector, however
several muons in the same sector are required to evaluate the associated
fit constants. There are of order $10^{5}$ sectors per region.

\section{FTK patterns banks}
\label{patbank}
\subsection{TSP bank of pattern candidates}
\label{TSPbank}

To prepare the pattern bank used for the pattern recognition, of
order $10^{11}$ pattern candidates are generated (about $10^9$ per region).
Each pattern candidate corresponds to eight coarse resolution hit
positions, associated with a track.
The eight hits correspond to the eight detector layers used for the
pattern recognition. 
The tracks are drawn from uniform distributions in the five track
parameters.
The corresponding hit positions, however, are not obtained from a full
simulation of the ATLAS tracker. 
Instead, the previously determined sectors and corresponding fit constants
are used to predict hit positions. 
When generating the $10^{11}$ pattern candidates, duplicates are
encountered. Only about one third of the patterns are unique, and 
duplicates with high multiplicity (coverage) are most 
important for the pattern recognition.
The full set of unique pattern candidates,
ordered by region, coverage and sector, is the so-called
thin-space-pattern bank (TSP bank). 
In the following, the algorithm to select the patterns from these
candidates for use with 
the AM boards is described. 

\subsection{AM pattern bank for use in the pattern recognition}
\label{AMbank}

The FTK hardware only supports a certain maximum number of
pattern. There are two AUX boards per region, corresponding to
$2^{24}$ ($16.8$ million) patterns per region.
As described above, three of the bits in each layer and each pattern are
ternary. By setting a ternary bit to the state $X$, the corresponding
pattern is valid for two hit numbers in the respective layer. By
setting a total number of $N_X$ bits across all eight layer to the state $X$,
the effective number of patterns stored in a single address increases
to $2^{N_X}$. The ternary bits thus can be used to vastly increase the number of
patterns stored in the AM chip. On the other hand, patterns with a
high number $N_X$ have a degraded spatial resolution and suffer from
an increased rate of combinatorial background (fakes).
A full set of $2^{24}$ patterns including ternary bits, determined for each of the
$64$ regions, is termed ``AM pattern bank''.

\subsection{Packing pattern candidates to the AM pattern bank}
\label{AMbankproduction}

For a single region, there are about $10^9$ pattern candidates. In this
section the basic algorithm is described to select the most relevant
patterns and pack them into the available $16.8$ million AM addresses,
also making use of the available ternary bits.

The primary sorting criterion is the pattern coverage. Pattern
candidates are ordered by their coverage, i.e. how often they have
been generated. The AM addresses are filled with pattern
candidates, starting with the highest coverage.
Once all addresses are filled, there is no room to include
further pattern candidates and the algorithm stops.

\begin{figure}[h]
	\centering
	\includegraphics[width=0.7\textwidth]{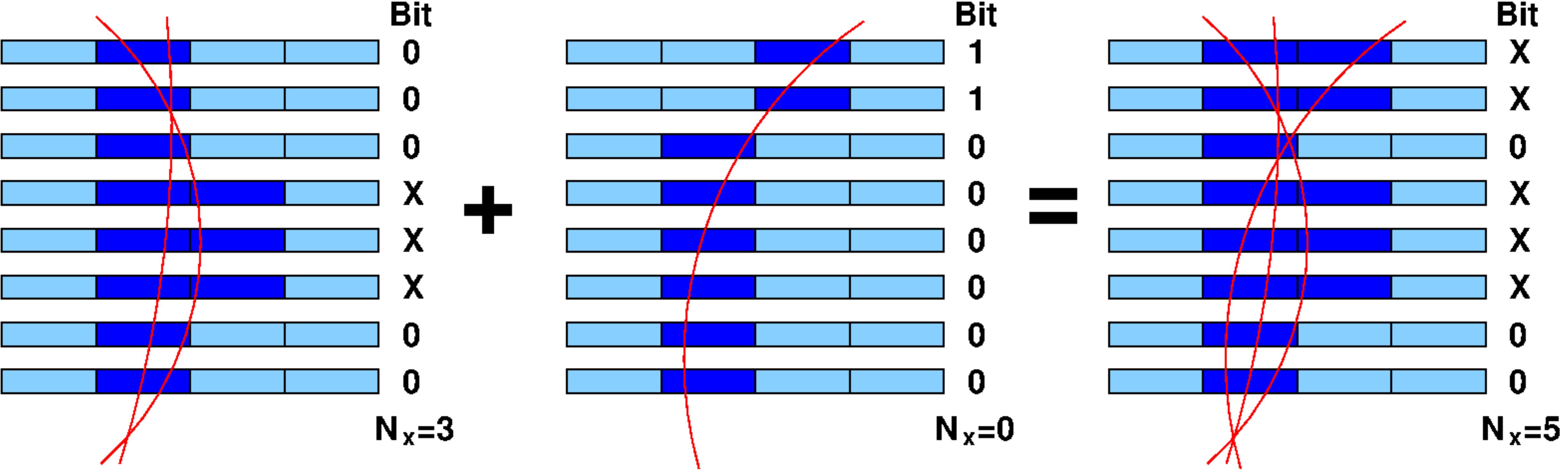}
	\caption{Example of a pattern with $N_{X}= 3$, changing to
          $N_{X}= 5$ when adding another pattern.}
	\label{fig:4}
\end{figure}

Using ternary bits, this algorithm is modified. The candidates are
still processed in order of decreasing coverage. However, where
possible, patterns are merged
with previously stored patterns, using the ternary bits. This is
illustrated in Figure \ref{fig:4}.
In the simplified example, shown for a single ternary bit per layer, the pattern
shown in the middle with least-significant bits ``00000011'' is merged
with a previously stored pattern ``00XXX000''. The resulting pattern
is ``00XXX0XX''.

In order to be able to tune the algorithm, the number of
ternary bits available for each layer can be chosen. In Figure
\ref{fig:4}, this number is set to one for each layer, while a maximum
of three ternary bits is possible in hardware.
A typical choice for the FTK system is to have one ternary bit per
coordinate, i.e.~two bits per pixel layer and one bit per SCT layer
\cite{Annovi2013}.
The disadvantage of this method is that some of the resulting patterns
have a rather high number of bits in state ``X'' and thus are
more susceptible to fakes.

Another method is to limit the number of ternary bits
across the whole pattern, $N_X$.
While, for example, the maximum number of three ternary bits is
allowed for each individual detector layer, the total number of
ternary bits, summed across all layers, is limited to a fixed
number, $N_X\le N_{X,\max}$. Choose, for example, $N_{X,\max}=4$. The
example shown in Figure \ref{fig:4} would lead to $N_X=5$. So in that
case the new pattern has to be stored at a new address, without
merging. It turns out that this algorithm is best in trading
efficiency against fake rate \cite{Stef2017}.

\section{WildCards optimisation}
\label{WC}

\begin{figure}
	\centering
	\includegraphics[width=0.375\textwidth]{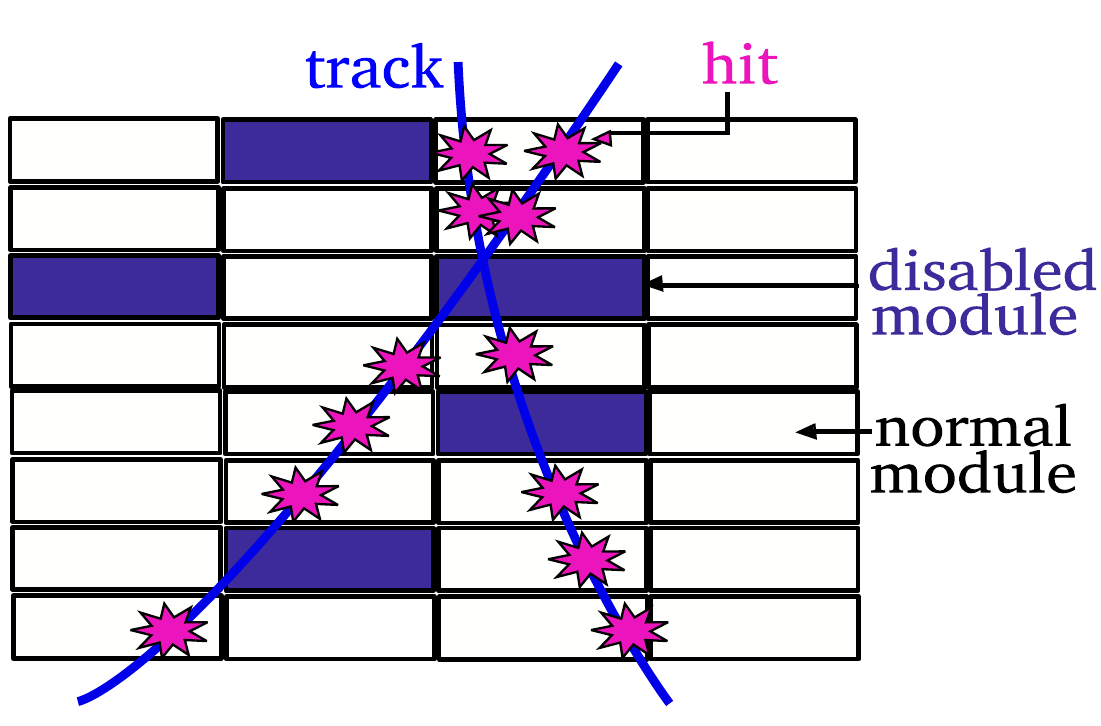} \includegraphics[width=0.3\textwidth]{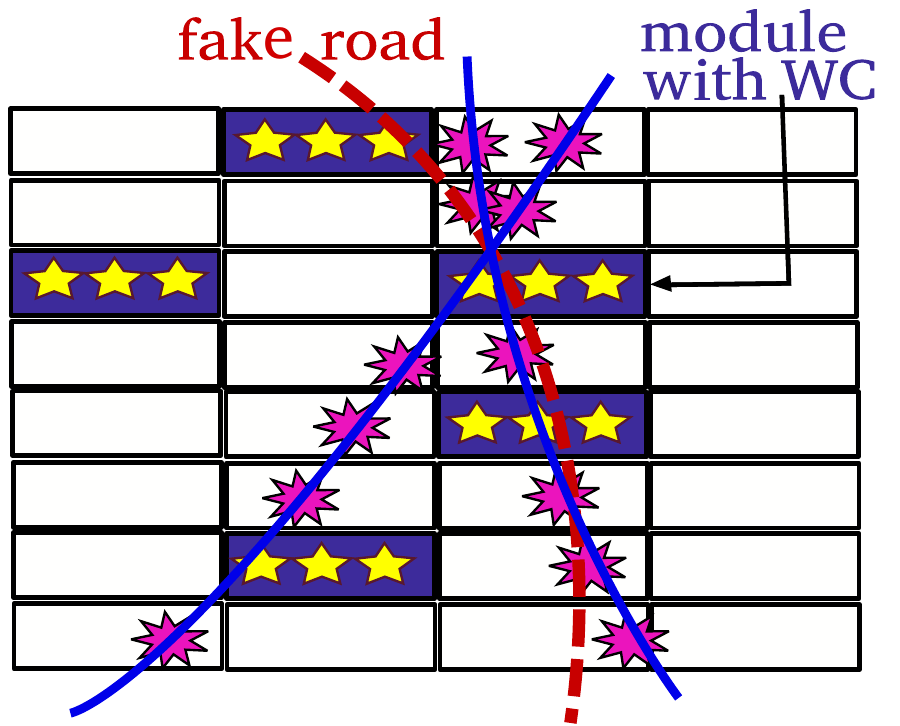}
	\caption{Illustration of tracks lost by disabled modules and their
          recovery by the wildcard algorithm. An example of
          combinatorical background (fake road) is also shown.}
	\label{fig:5}
\end{figure}

To find a track in the FTK system, a pattern match is required in 7 or
8 layers.
Owing to inevitable hardware failures, there are disabled modules, thus
creating inefficiencies. Examples are shown in Figure
\ref{fig:5} (left).
The blue tracks cannot be reconstructed, because of disabled 
modules on the trajectory.
Using realistic configurations of disabled hardware, FTK efficiency losses of order
$2-4\%$ are observed in parts of the detector.

\vskip-20pt
\subsection{WildCards Algorithm}
\label{WCalgo}

To recover the efficiency losses caused by disabled modules, a
WildCard (WC) algorithm is used.
In this algorithm, patterns which require hits on disabled module are
modified such that they are recognized already if six out of the
remaining seven possible hits are found.
Figure \ref{fig:5} (right) illustrates the effect of the WC algorithm.
The blue tracks are recovered.
However, in addition a ``fake road'', shown in red, is also found from
combinatorial background.
In other words, the WC algorithm does recover the efficiency but at
the same time can lead to a much increased number of fake
roads that endanger operation of the system due to congestion effects.
The primary measure to control these effects is to limit the number of
wildcards per pattern to one.
In the unlikely case that there are two disabled modules for a given
pattern, only one of them will have a wildcard set.

\subsection{The wildcard penalty algorithm}
\label{WCpen}

To gain flexibility over the amount of fakes in the presence of wildcards, a WC
penalty algorithm is introduced when selecting the pattern candidates
for the AM pattern bank.
The algorithm is designed such that it avoids to have wildcards and a
large number of ternary bits in the same patterns.
The original condition $N_{X}\le N_{X,\max}$ is modified by adding a
penalty term
\begin{equation}
 N_{X} + P\cdot N_{WC} \leq N_{X,\max{}},
\end{equation}
where $P$ is the penalty and $N_{WC}$ is the number of wildcards in the
pattern (i.e. zero or unity).
The penalty forces patterns on which the WC algorithm is applied to
have a lower maximum number of ternary bits in state "X".

\subsection{Comparisons}
\label{meth}

To validate the performance and optimize the WC algorithm and WC
penalty, different configurations are compared using the FTK simulation \cite{JAA2015}.
In this study the maximum number of ternary bits ($N_{X,\max{}}$) is
set to seven for barrel regions ($0<\vert\eta\vert\lesssim1.4$) and to
four for end-cap regions  ($1.4\lesssim\vert\eta\vert$).
Five sets of pattern banks are investigated:
\begin{itemize}
\item four banks produced with wildcards and $P={0,1,2,3}$, respectively;
\item a ``nominal'' bank produced without disabled modules or wildcards,
  corresponding to the ideal detector configuration.
\end{itemize}
When running the simulation with the nominal bank, there 
is the option to enable wildcards: list of disabled modules are defined with wildcards switched on.
The result is expected to be similar to the case $P=0$, although
patterns can be  packed more efficiently when knowing about disabled
modules during bank production.
Another option which can be tested with the nominal bank is to
simply disable the modules, without using wildcards.

\section{RESULTS}
\label{results}

\begin{figure}[h]
  \begin{minipage}[t]{0.48\textwidth}
  \centering
  \includegraphics[width=0.98\textwidth]{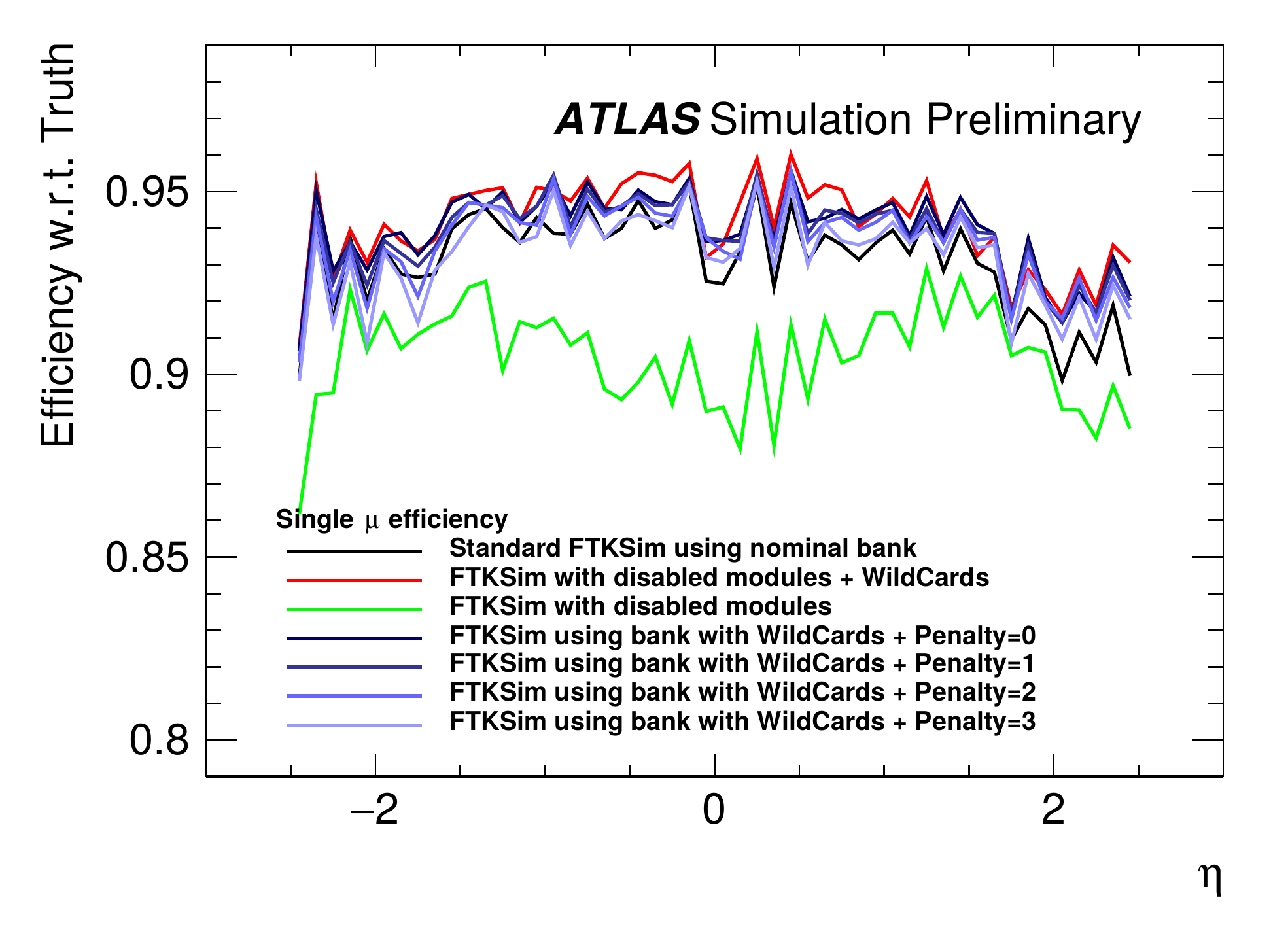} 
  \vskip-10pt
  \caption{
      \label{fig:7}
      Efficiency for single particles as determined
      with different pattern bank configurations
      as a function of the particle pseudorapidity. }
  \end{minipage}
  \hfill
  \begin{minipage}[t]{0.48\textwidth}
    \includegraphics[width=0.98\textwidth]{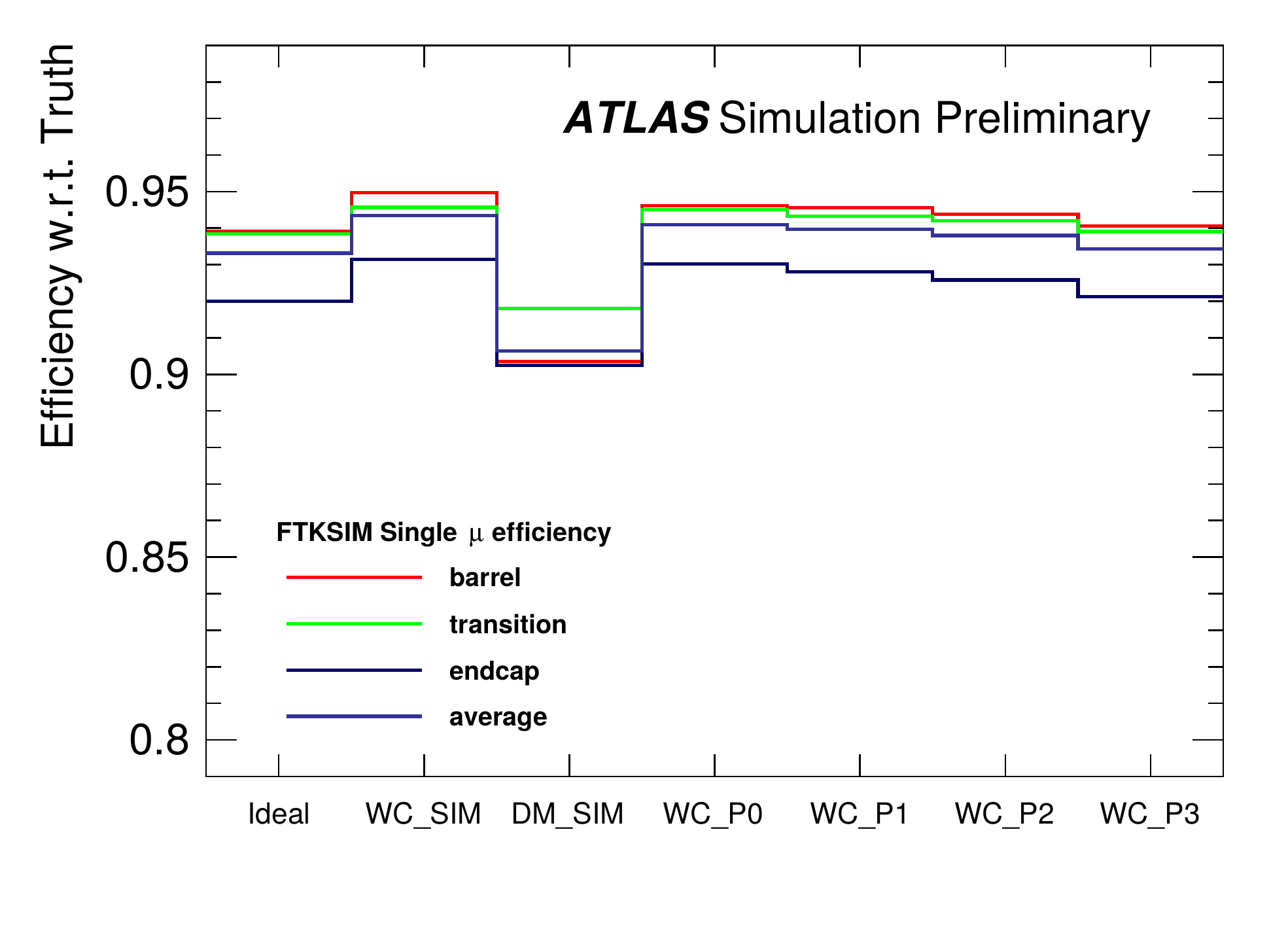}
  \vskip-10pt
    \caption{    \label{fig:8}
      Average efficiency for single
      particles determined with different
      pattern bank configurations in four polar angular
      ranges as described in the text.}
  \end{minipage} 
\end{figure}

In Figure \ref{fig:7}, efficiencies as a function of
the muon pseudorapidity $\eta$  are compared for the seven
configurations presented in section \ref{meth}.
The black line shows the simulation output assuming perfect detector
conditions.
When simulating disabled modules (green line), sizable inefficiencies
are observed, in particular in those parts of the detector which suffer from
the highest number of disabled modules.
The efficiency improves significantly when using the WC algorithm (red
line).
The effect of producing dedicated pattern banks with wildcard and a
penalty term in the range $0$-$3$ is also shown.
Increasing penalties reduce somewhat the efficiency, but even with the
highest penalty studied, the efficiency is close to the ideal
detector and superior to the case of disabled modules.
In Figure \ref{fig:8}, efficiencies are studied for the
seven configurations, where averages are taken over ranges of
polar angle.
The configurations are labelled as: four settings of the penalty
``WC\_P0'',``WC\_P1'',``WC\_P2'',``WC\_P3''; nominal bank with
ideal detector ``Ideal''; nominal bank with WC ``WC\_SIM'', nominal
bank with disabled modules: ``DM\_SIM''.
The ranges studies are: barrel $\vert\eta\vert<1.1$,
transition $1.1<\vert\eta\vert<1.6$, endcap $\vert\eta\vert>1.6$.
The average over the full polar angular range is also shown.
The overall efficiency is lowest in the endcap.
The relative impact of dead modules is largest in the barrel,
where the majority of the disabled modules are located.
As expected,
the nominal bank with wildcards and the simulation with $P=0$ have
similar efficiencies.

\begin{figure}[h]
  \includegraphics[width=15.5pc]{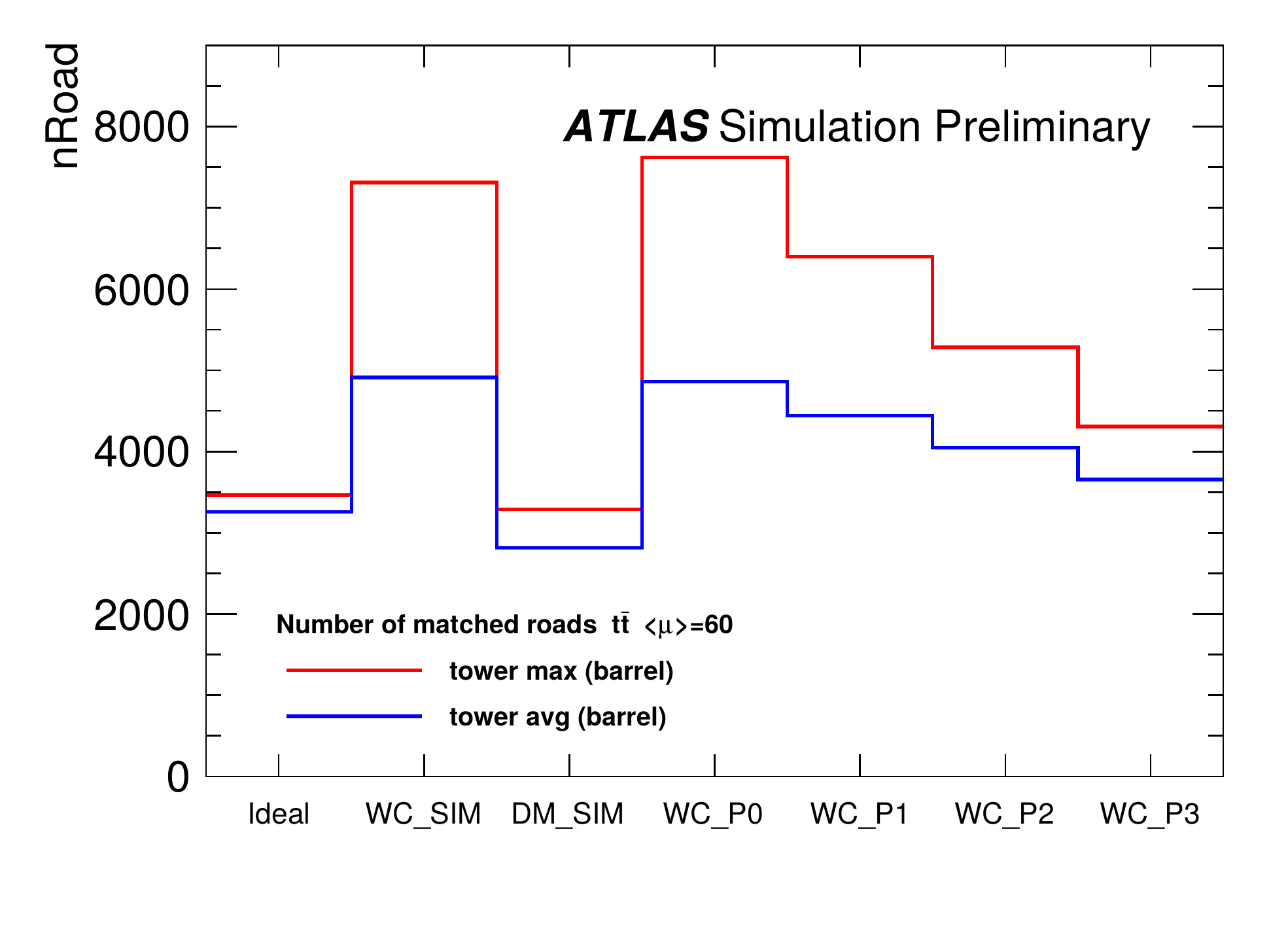}
  \includegraphics[width=15.5pc]{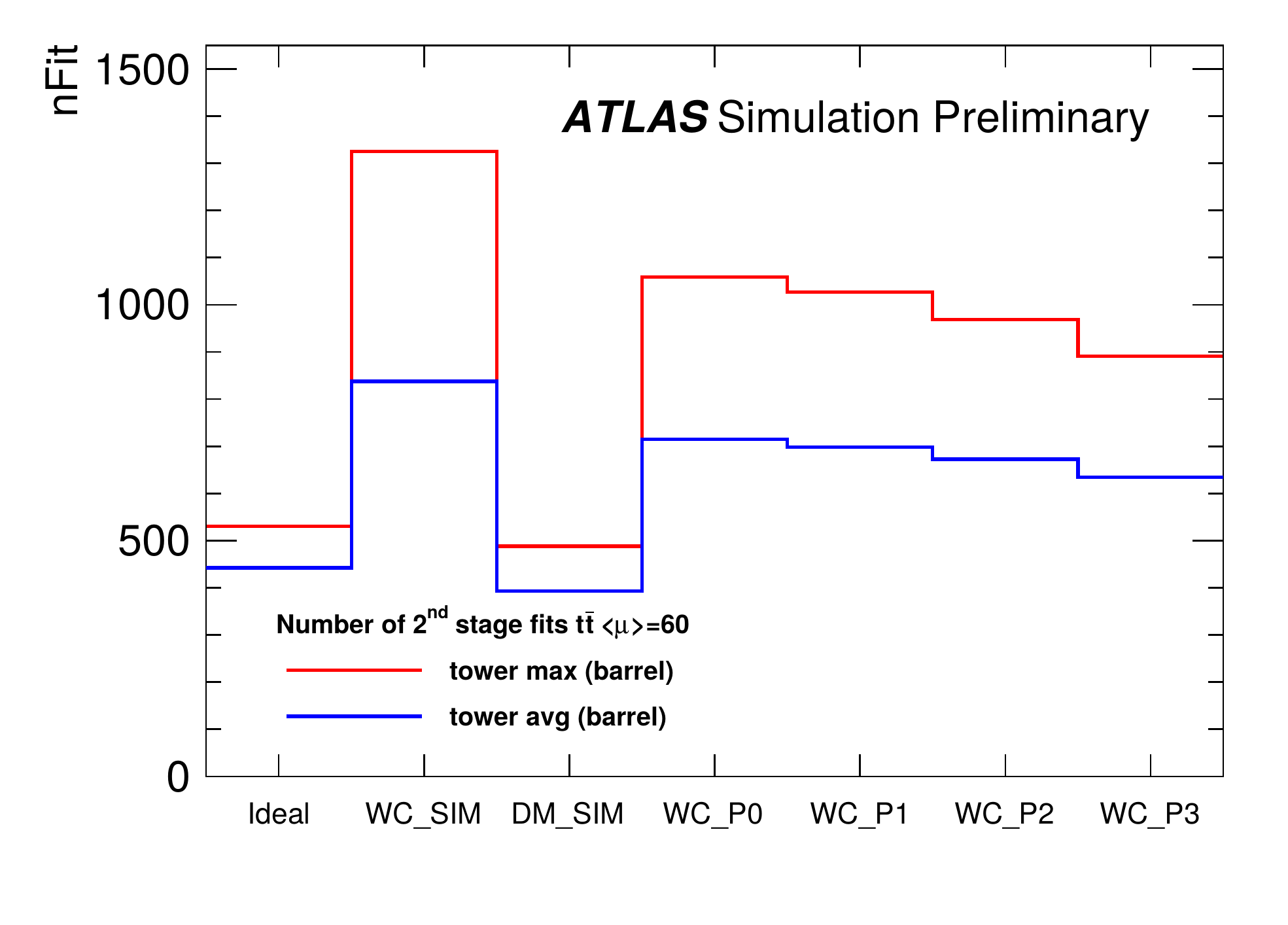}
  \vskip-20pt
  \caption{Number of roads (left) and number of second stage track
    fits (right) per event and per FTK
    region as determined with different pattern bank
    configurations. The numbers are
    averaged over all events taken from a
    simulation of $t\bar{t}$ decays with a pileup of 60
    minimum bias events. In each subfigure, the two
    curves correspond to 
    the average over all barrel regions and to the barrel
    region with the largest event average, respectively. }
  \label{fig:9}
\end{figure}

Figure \ref{fig:9} shows the number of roads (left) and the number of
track fits (right), averaged over all events of a simulation of  $t\bar{t}$ decays at a
pileup of $60$.
The number of roads and the number of second stage track fits both serve as typical
``dataflow'' quantities, to quantify the danger of congestion effects in the 
FTK system. 
When comparing the nominal bank with and without dead modules
``Ideal'' and ``DM\_SIM'', there is only a small change in the dataflow.
There is, however, a clear effect of the use of WC on the dataflow
quantities. Switching on wildcards for the nominal bank causes the
dataflow quantities to increase dramatically. When using dedicated
pattern banks with penalties ranging from $0$ to $3$, the increase can
be controlled, although an increase in dataflow is present even for
the highest penalty investigated.

\section{CONCLUSION}
\label{conclusion}

The FTK is a hardware-based track finding system to be operated with
the ATLAS trigger. It reconstructs all tracks of an event with transverse
momentum $p_T>1$~GeV at a rate of $100$~kHz and a latency of
$100\,\mu s$ for use in early stages of the high-level trigger decision.

Pattern recognition in the FTK is based on a predefined pattern
bank. Only those tracks are reconstructed which have hits
corresponding to seven out of eight possible layers in one of the
predefined patterns. The patterns can be tuned in size by means of
ternary logic.

Disabled modules are inevitable when operating a detector in the LHC
environment. Using the nominal FTK pattern
recognition algorithm, disabled modules cause track reconstruction
inefficiencies of order $2$-$4\%$.

Wildcard algorithms are investigated in this article.  When simply
removing the disabled layers from the pattern recognition, the
efficiency can be recovered, however at the cost of increasing the
rate of combinatorial background. This effect endangers the operation
of the FTK, which can only handle a certain average number of extra patterns
or track fits per event. An algorithm is developed which dynamically
reduces the pattern size in those parts of the pattern bank where
wildcards are used. Simulations show that the algorithm can be useful
to find a good compromise between excessive combinatorial background
and the best possible efficiency.


\end{document}